\documentclass[12pt]{elsarticle}

\usepackage[a4paper,top=1in,bottom=1in]{geometry}
\usepackage{amsmath}
\usepackage{hyperref}
\usepackage{graphicx}
\usepackage{color}

\journal{Economics of Education Review}

\makeatletter
\def\ps@pprintTitle{%
  \let\@oddhead\@empty
  \let\@evenhead\@empty
  \def\@oddfoot{\reset@font\hfil\thepage\hfil}
  \let\@evenfoot\@oddfoot
}
\makeatother

\date{March 12, 2016}
\title{Reducing the role of random numbers in matching algorithms for school admission}
\author{Wouter Hulsbergen} 
\address{Nikhef National Institute for Subatomic Physics, Amsterdam, The Netherlands}
\ead{wouter.hulsbergen@nikhef.nl}

\begin{document}

\begin{frontmatter}



\begin{abstract}
  New methods for solving the college admissions problem with
  indifference are presented and characterised with a Monte Carlo
  simulation in a variety of simple scenarios. Based on a qualifier
  defined as the average rank, it is found that these methods are more
  efficient than the Boston and Deferred Acceptance algorithms. The
  improvement in efficiency is directly related to the reduced role of
  random tie-breakers. The strategy-proofness of the new methods is
  assessed as well.
\end{abstract}


\end{frontmatter}

\section{Introduction}

After six years of primary school education pupils in the Netherlands
choose a secondary school. In Amsterdam children have abundant
choice, with up to a dozen different schools at each of the available
school levels. Pupils will have a preference for a certain school
based primarily on the distance to their home, objective and less
objective public data on the quality of education, and the
impression the schools make in their advertisement and `open days'.

Unfortunately, the number of pupils a school can accept is not
necessarily proportional to its popularity. Therefore, the local
government has introduced a matching system to assign pupils to
schools, similar to that used in many other cities in the
world.\footnote{In Amsterdam the matching system is referred to with
  the Dutch word \emph{kernprocedure}.} Each pupil composes an ordered
list of schools of their choice. \footnote{In the language of the
  field we say that students only provide their \emph{ordinal}
  preferences. Their \emph{cardinal} preferences, how much they value
  schools compared to one another on a more continuous scale, are
  unknown to the matching system.} Based on the collection of these
preference lists the matching is performed. Although the pupils have
clear preferences, the schools in the Amsterdam system are not allowed
to differentiate pupils. In literature, this matching problem is
also referred to as the \emph{college admissions problem with
  indifference}~\cite{Roth:1989,Roth:2007}. As all pupils may hand in exactly
the same preference list, the system requires a method for
arbitration, or \emph{tie-breaking}, at popular schools. This
arbitration is performed by assigning pupils a lottery number.

The matching system relies on a procedure, or \emph{algorithm}, to
turn the available list of choices and lottery numbers into a final
assignment of pupils to schools. Different methods were applied
in the past. Up to the year 2014 the city of Amsterdam effectively
used the so-called ``Boston'' algorithm. A disadvantage of the Boston
method is that it is not
\emph{strategy-proof}~\cite{Abdulkadiroglu:2003}, a well-known concept
in game theory: Using predictions on how other pupils will vote
(e.g. from the popularity of schools in previous years), pupils may
benefit from providing a preference list that is different from their
true preference list.

Whether this is actually a disadvantage or not is a topic for
debate~\cite{Abdulkadiroglu:2008,Abdulkadiroglu:2009}, but nevertheless, in 2015 the
Deferred Acceptance (DA) algorithm~\cite{Gale:1962} was introduced,
which is known to be strategy-proof~\cite{Dubins:1981}. In the
implementation chosen in Amsterdam, a different random tie-breaker for
each school was used. This algorithms is sometimes abbreviated as
\emph{DA-MTB}~\cite{Oosterbeek:2015}, where MTB stands for multiple
tie-breakers. Unfortunately, the DA-MTB algorithm is not what is
called \emph{Pareto efficient}: two students may find that they both
end up higher on their preference list if they exchange schools after
the assignment. Not surprisingly this led to a public outcry from
parents that had not anticipated this. As a result the system was yet
again changed for the current calendar year: In 2016 Amsterdam will
apply the DA algorithm with a single random tie-breaker
(a.k.a. \emph{DA-STB}).

The college admission problem is a well-studied problem. Yet, the pace
with which algorithms are replaced in Amsterdam illustrates the lack
of consensus on how to decide what is the best algorithm. The aim of
this article is threefold. First, we introduce a qualifier, a single
number that measure for the welfare of the students, by which one can
compare algorithms. Second, we introduce alternatives for the Boston
and DA algorithms. One alternative, which we will call the `Zeeburg
algorithm', is a matching algorithm specifically designed to minimize
the number of comparisons made with the random tie-breaker. The other
alternative is a method to improve on the solutions given by DA and
Boston by introducing pairwise exchanges.  Using simulated data for a
number of different scenarios we show that these algorithms are less
sensitive to the results of the lottery and better respect the
preferences of the pupils. Finally, we argue that although the new
algorithms are not strategy-proof, there is a compelling reason for
favouring them over the algorithm that is currently applied in
Amsterdam: even students that do not apply a strategy are better off.

\section{The average rank as a welfare qualifier}

Before discussing the algorithms and results, we briefly introduce
notation and a few definitions.  We consider a set of $M$ schools
labeled by an index $j$. Every school has place for $N_j$ new
pupils. We label the pupils by an index $i$ and assume that the total
number of pupils $N$ is smaller or equal to the total number of
places, $ N \leq \sum_j N_j$.

Based on personel preference, every pupil ranks the $M$ available
schools in a list. We call that ranked list a \emph{preference} and
denote it with the symbol $p$. For example, given four schools with
labels 1, 2, 3 and 4, the list of pupil $i$, could look like
\begin{equation*}
  p_i \; = \; ( 3,1,2,4 )
\end{equation*}
such that each of the four schools appears exactly once. We label the
$j$-th entry in the list by $p_{i,j}$ such that the most preferred
school is $p_{i,1}$ (school $3$ in the example) and the least
preferred school is $p_{i,N}$.  The total set of preferences $\{ p_i
\}$ of all students is the \emph{preference set}, or simply dataset.

The result of the matching procedure is an assignment of pupils to
schools $j$. We denote the value of $j$ for student $i$ by the
$a_{i}$. Every pupil is assigned to only one school and that
school is $a_i$. Furthermore, every school can be assigned as most
as many pupils as it has place, or, for every school $j$,
\begin{equation}
  \sum_{\text{pupils $i$}} \delta_{a_{i},j} \; \leq \; N_j
\end{equation}
where $\delta_{ij}$ is $1$ for $i=j$ and $0$ otherwise. We call a set
of $N$ assignments $ \{ a_{i} \} $ a \emph{solution} $S$ if it
satisfies this property.

We now introduce a simple qualifier to be able to rank solutions. The
lower the assigned school ranks on the pupils preference list, the
less satisfied the pupil is with the assignment. For a given solution
$S$ we quantify the dissatisfaction as the pupil's rank of the
assigned school, or
\begin{equation}
  r^S_i = \text{``value of $j$ for which $p_{i,j} = a^S_i$''} 
\end{equation}
For instance, in the example above, the pupil's rank for a solution in
which it would get assigned school~$1$ is two, etc. Our welfare
qualifier for the solution $S$ is now simply the average rank
\begin{equation}
  Q(S) \; = \; \frac{1}{N} \sum_i r^S_i .
\end{equation}
We define the optimal solution as the solution for which $Q$ is
minimal. 

The optimal solution is not necessarily unique: there may be several
solutions with the same value $Q$. In theory these solutions could be
found by simply trying all possible assignments. Unfortunately, in any
realistic scenario the number of possible combinations is far too
large to try even a small fraction of them.\footnote{With $N$ pupils
  divided equally over $M$ schools the total number of permutations
  equals $N!  / (N/M)^{M}$. With 1000 pupils and 10 schools there are
  more permutations than atoms in the universe! CHECK} Therefore, in
practice it is not easy to find even a single optimal solution.

As we shall see below the Boston algorithm maximizes the number of
students with rank one. One may wonder if the optimal solution always
satisfies this property as well. However, a simple counter example
shows that it does not.\footnote{Take four schools with one student
  each. If the students preference sets are ${ (1,3,2,4), (2,1,3,4),
    (3,4,1,2),(2,3,1,4) }$, the solution with the most pupils at rank
  one is $a = (1,2,3,4)$, which has $Q = 7/4$. The solution with the
  smallest average rank is $a = (1,2,4,3)$, with $Q=6/4$.}  This
illustrates that the solution cannot be found by first
maximising the number of rank one assignments, then rank two, etc. As
far as we know, there is no algorithm that can find the optimal
solution to this matching problem in a reasonable amount of time.

We label an algorithm \emph{efficient} if it provides the optimal
solution. Lacking a truly efficient algorithm, all that we can wish
for is an algorithms that is \emph{nearly} efficient. In the
following, we shall call one algorithm more efficient than another
algorithm if on an ensemble of similar datasets it gives a solution
that on average has a smaller value for $Q$.

\section{Characterisation of  algorithms in a Monte Carlo simulation}

The DA algorithm was originally developed to solve a two-sided market
problem in which both sides have a strict ordinal preference to
partners on the other side~\cite{Gale:1962}.  The college matching in
Amsterdam and many other cities is different: Schools are not allowed
to rank students. To apply the traditional algorithms anyway, a
sequence of random numbers, the \emph{tie-breaker}, takes the role of
the ordinal preferences of the schools.

The Boston and DA algorithms can use a random tie-breaker in two
ways~\cite{Abdulkadiroglu:2008}: If all schools share the same
tie-breaker, we talk about the \emph{single tie-breaker} (STB)
variant. If every school has its own tie-breaker, we call it the
\emph{multiple tie-breaker} (MTB) variant. It can be shown that,
independent of the tie-breaker, the Boston algorithm is not
strategy-proof, while the DA-MTB algorithm is not Pareto efficient.

Using random numbers as a tie-breaker may lead to inefficient
matching, because the real preferences of the students for schools
compete with the random preference from schools for
students~\cite{Abdulkadiroglu:2008}. To illustrate this we now compare
the behaviour of these algorithms in a Monte Carlo simulation.

\subsection{Description of the simulation}

In~\cite{Oosterbeek:2015} the matching algorithms are compared on an
actual dataset collected in the year 2015 in Amsterdam. This has the
advantage that it corresponds to a real scenario, with real
preferences of students including correlations. It has the
disadvantage that only a single dataset can be used for the
comparison: It tells little about the sensitivity to variations in the
input dataset. One could imagine that an algorithm can be tuned to be
efficient on one dataset, but behaves differently on the
next.\footnote{The authors of~\cite{Oosterbeek:2015} have used their
  single input set to simulate multiple experiments by varying the
  random numbers for the tie-breakers. This helps to understand the
  sensitivity to the tie-breakers, but not to variations in the data,
  \emph{e.g.} variations in students in different years.}

As an alternative we have chosen to define a set of simple scenarios
that allow us to randomly generate datasets. This technique is called
a Monte Carlo simulation. One generated dataset is called an
\emph{experiment}. We consider a matching problem with 10 schools,
labeled 1 to 10. Each school has place for 100 pupils, giving a
total of 1000 pupils. Consequently, a single generated dataset
consists of 1000 rankings of ten schools.

We now consider four scenarios that differ in the way students
\emph{on average} fill in their preference list. More specifically, in
each scenario we choose how often a pupil puts a particular school as
its first choice. The selected scenarios are given in
table~\ref{tab:scenarios}.

\begin{table}[htb]
  \centerline{
    \small
    \begin{tabular}{|c||c|c|c|c c|}
      \hline
      {} & scenario A & scenario B & scenario C & \multicolumn{2}{c|}{scenario D} \\
      \hline
      fraction of students & 100\% & 100\% & 100\% & 60\% & 40\% \\
      \hline
      school 1  & 1 & 10 & 50 & 20 & 1  \\
      school 2  & 1 &  9 & 50 & 20 & 1  \\
      school 3  & 1 &  8 & 10  & 20 & 1  \\
      school 4  & 1 &  7 & 10  & 20 & 1  \\
      school 5  & 1 &  6 & 10  & 20 & 1  \\
      school 6  & 1 &  5 & 10  &  1 & 20 \\
      school 7  & 1 &  4 & 10  &  1 & 20 \\
      school 8  & 1 &  3 & 10  &  1 & 20 \\ 
      school 9  & 1 &  2 & 1  &  1 & 20 \\
      school 10 & 1 &  1 & 1  &  1 & 20 \\
      \hline
    \end{tabular}}
  \caption{Relative popularity of schools in the different
    scenarios. A value of 10 means that a school is ten times more
    likely to appear as a first choice than a school with a value of
    1. In scenario D two populations of students are simulated, with
    a relative size of 60:40.}
  \label{tab:scenarios}
\end{table}

In scenario A all schools are equally popular. In scenario B school~1
is ten times more popular than school 10, and the rest is in between at fixed
intervals. Scenario C is a variation of this, with two
highly popular schools and two highly unpopular schools. We shall use
this scenario to assess the effect of strategic ranking. The
motivation for scenario D is given later.

\begin{figure}[htb]
  \centerline{\includegraphics[width=\textwidth]{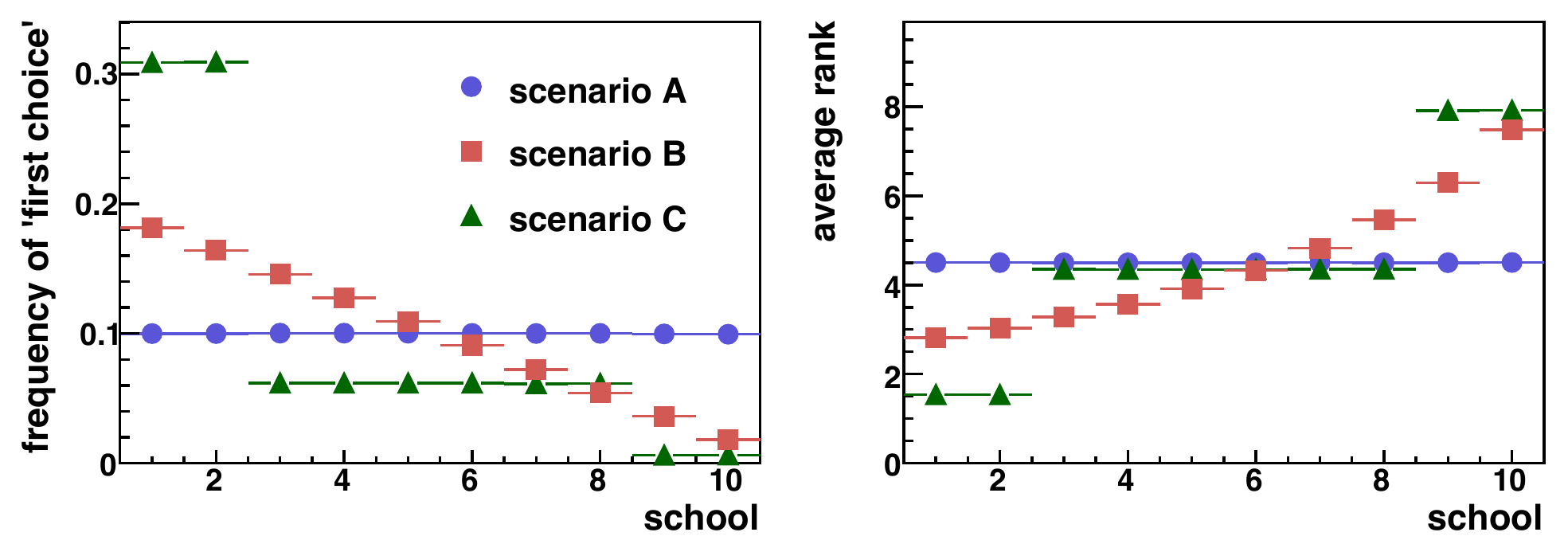}}
  \caption{Fraction of times a school is ranked first (left) and
    average rank (right) as function of school number in scenarios
    A, B and C, measured over 100 experiments.}
  \label{fig:generatedschoolrank}
\end{figure}

A single dataset is now generated as follows. Using the popularity of
the schools a set of random numbers determines the order of the
schools for each pupil. To determine the first school, the relative
popularities are normalized to add up to one and a random number in
the interval $[0,1)$ is thrown. The quantile of the random number
determines the first school. A second random number determines the
second school on the list, by considering the popularity normalized
over the remaining schools. This procedure is repeated, until there
are no schools left. It is then performed a thousand times --- once
for each pupil --- to obtain the dataset for one
experiment. Figure~\ref{fig:generatedschoolrank} shows for scenarios
A, B and C how often a particular school appears first, and which
position it takes on average on a pupil's preference list, averaged
over 1000 experiments.

In scenarios A, B and C the preferences of the pupils are
uncorrelated: The selection of the preference for the second school is
independent of which school was put first.  In practice, pupil
preferences are often correlated, for example because students prefer
schools that are close to their neighbourhood. To include also a
scenario with correlations we consider yet another scenario, labeled
by D in the table. In this scenario there are two categories of
students who have each their own popularity assignment: the first
set of students strongly prefer the first five schools while the
second set prefer the last five schools. If the two categories had
an equal number of students, we could just factorise the matching
problem, and effectively end up with scenario A. However, to simulate
also the effect of an imbalance in capacity, we generate 60\% of the
students in the first category and 40\% in the second category. That
means that for 20\% of the students in the first category no school in
their top-five can be assigned.

\subsection{Results of the simulation}

Given a simulated dataset we can now use the algorithms discussed
above to obtain a matching. We have found that the Boston-MTB and
Boston-STB are close in behaviour on all considered scenarios, and
therefore we only consider Boston with a single tie-breaker.

Besides DA and Boston we also test a new algorithm that we have called
the Zeeburg algorithm. The details are described in
appendix~\ref{app:zeeburg}. In brief, this algorithm minimizes the
number of times the tie-breaker is used to compare students by making
students jump to a queue of a school that appears later in their
preference list if by doing so they are guaranteed to be admitted to
that school. In some sense, the algorithm encodes a strategy for the
students. The Zeeburg algorithm is Pareto efficient and
stable\footnote{In a two-sided market problem with strict preferences
  on both sides a stable solution is a solution in which there is no
  pair of a student and a school that would prefer each other over
  their actual assigned partner(s). However, since the school
  preferences in the college problem with indifference are entirely
  fictitious, stability is not really relevant here.}, but not
strategy-proof.

\begin{figure}[htb]
  \centerline{\includegraphics[width=1.0\textwidth]{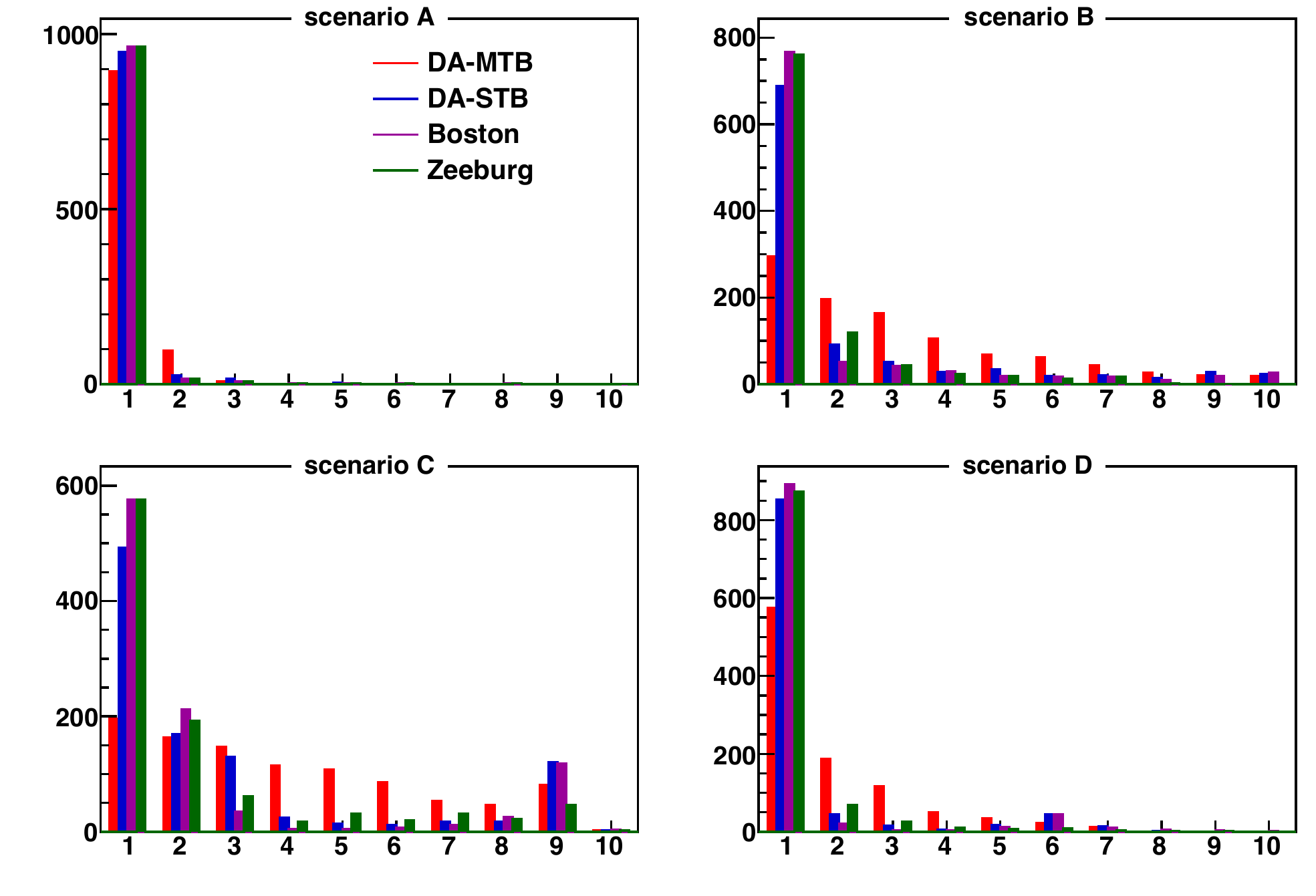}}
  \caption{Distribution of rank for one experiment in each of the four
    scenarios sketched in the text for the Boston, DA-STB and DA-MTB
    algorithms. Each distribution has a thousand entries,
    corresponding to the thousand students in the experiment.}
  \label{fig:oneexperiment}
\end{figure}

Figure~\ref{fig:oneexperiment} shows the distribution of the assigned
rank for one single experiment for each of the four scenarios. The
first bin shows how many students get assigned to the school of their
first choice, the second bin their second choice, etc. These
distributions will look different for every experiment, because the
preferences of the students differ and because the tie-breakers used in
the matching algorithms differ.

\begin{figure}[htb]
  \centerline{\includegraphics[width=\textwidth]{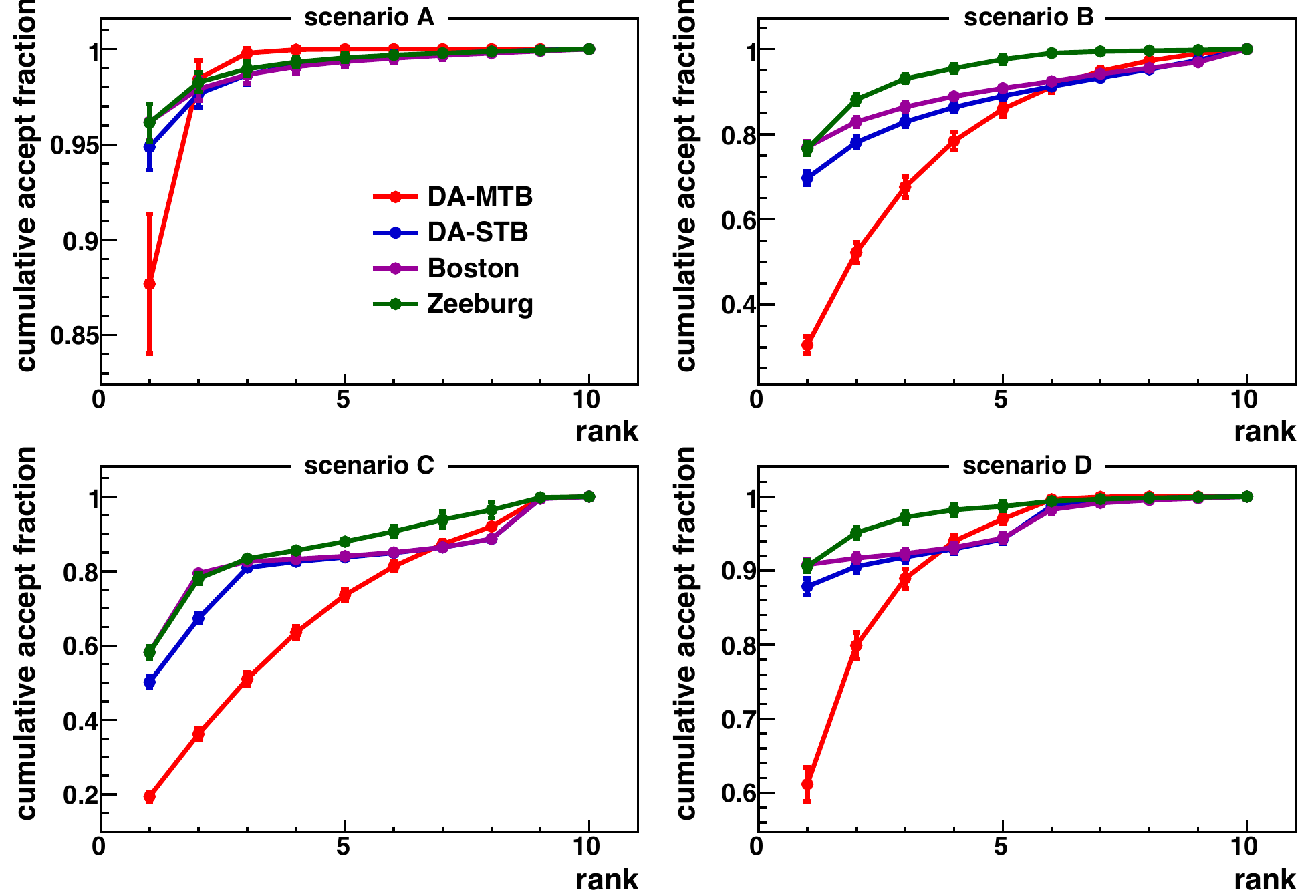}}
  \caption{Average cumulative acceptance functions for the four
    considered scenarios and for the Boston, DA-STB and DA-MTB
    algorithms, averaged over 1000 experiments. The vertical error
    bars correspond to the standard deviation of the variation between
    experiments.}
  \label{fig:acceptancecurves}
\end{figure}

A convenient way to summarise the information in
Fig.~\ref{fig:oneexperiment} for many experiments is to integrate this
distribution, normalize it and average over the experiments. The
result of this is shown in Fig~\ref{fig:acceptancecurves}. The curves
in this figure show which fraction of students get assigned to the
school of their first choice, to their first \emph{or} second choice,
etc.

It is clear from these graphs that independent of the scenario, the
Boston algorithms assigns most pupils to the school of their first
choice. This is a property of the algorithm: it actually assigns the
maximum possible number of students to their first choice. The DA-MTB
scores poorly when it comes to the first choice, but it has a smaller
tail. The reason for this is that with multiple tie-breakers, it is
unlikely that a student is unlucky at every school.

\begin{figure}[htb]
  \centerline{\includegraphics[width=\textwidth]{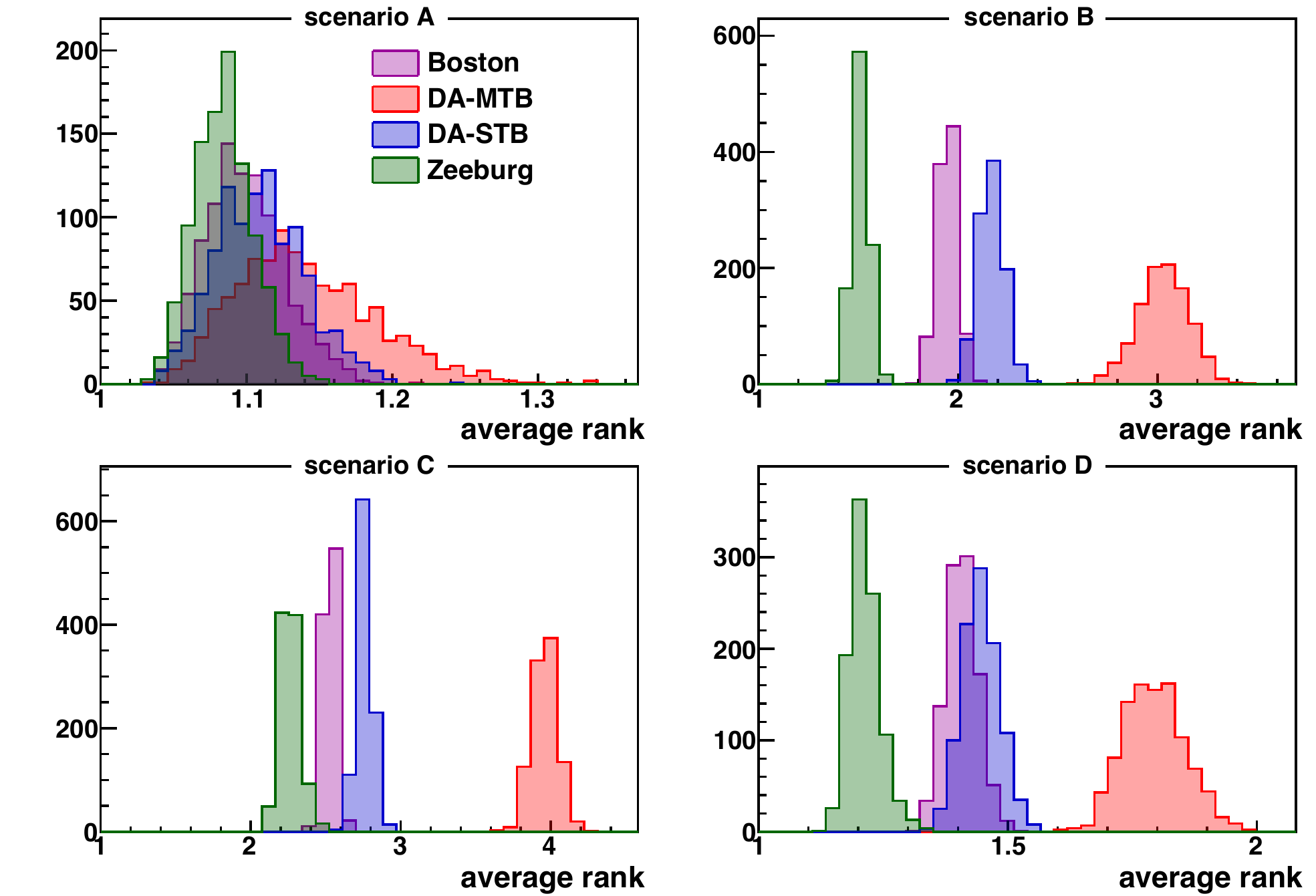}}
  \caption{Distribution of the average rank $Q$ over 1000 experiments
    for the four different scenarios in the Boston, DA-STB, DA-MTB and Zeeburg
    algorithms.}
  \label{fig:rankdistribution}
\end{figure}

Each of the points in Fig.~\ref{fig:acceptancecurves} has a vertical
`error' bar. The size of the error reflects the variation in the
integrals between the different experiments. This variation is larger
for the DA-MTB algorithms than for the Boston algorithm, because the
former is more sensitive to the random numbers in the tie-breaker.
Another way to represent the variation between experiments is to
consider the distribution of the average rank (our qualifier $Q$) in
each experiment, shown in Fig.~\ref{fig:rankdistribution}. In scenario
A the difference between the algorithms is small, but in all others it
is substantial. In terms of the qualifier defined above, there is
clear order in the efficiency of the four algorithms, with Zeeburg
having the highest efficiency and DA-MTB the lowest. For example, in
scenario B students are on average assigned to their third rank
school by the DA-MTB algorithm, while the average assignment of
Zeeburg is between the first and second ranked school.

\begin{figure}[htb]
  \centerline{\includegraphics[width=\textwidth]{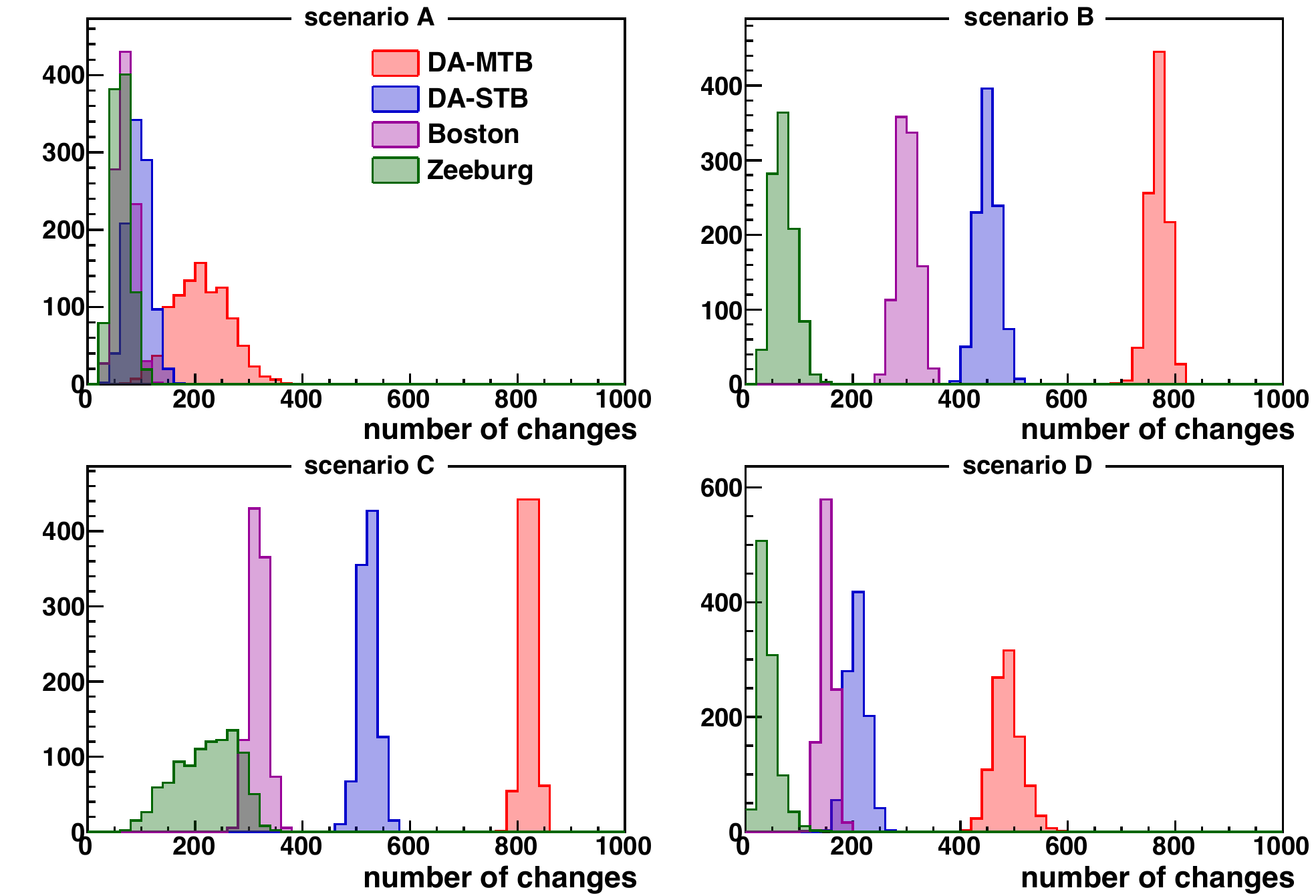}}
  \caption{Number of students that gets a different assignment in two
    consecutive calls to the same algorithm in 1000 experiments.}
  \label{fig:schoolswaps}
\end{figure}

Figure~\ref{fig:rankdistribution} also indicates a large variation in
the rank of an algorithm between different experiments. This variation
has two sources, namely the actual differences in the datasets and the
random character of the tie-breakers. To illustrate the importance of
the latter we show yet another distribution. For each experiment we
run the algorithms a second time, but with a different tie-breaker, a
different student lottery. For each algorithm we now count how many
students are the second time assigned to a different school, that is,
how `deterministic' the algorithm is. The result is shown in
Fig.~\ref{fig:schoolswaps} for all four algorithms. Comparing to
Fig.~\ref{fig:rankdistribution} we note that that the sensitivity to
the tie-breaker is correlated with the efficiency: the less important
the tie-breaker, the more efficient the algorithm.

\subsection{The pairwise exchange method}

Given a particular set of matches one can improve the average ranking
using \emph{pairwise exchanges} (PE), a swap of the schools assigned
to a pair of students. In principle, using pairwise exchanges one can
transform any solution into any other, including the optimal
solution. In practice, in order to limit the time-consumption of such
an algorithm, it is necessary to limit the set of considered
exchanges.

In~\cite{Oosterbeek:2015} only pairwise exchanges that improved the
ranking for both pupils involves in the swap were considered.  If any
such swaps can be found, the original solution was not Pareto
efficient.  However, exchanges that reduce the sum of the ranks of the
two pupils improve the solution as well. Therefore, in order for the
pairwise exchange method to be effective, such exchanges should be
considered. Besides pairwise exchanges, one can also consider
exchanges of higher order in which the average
improves. Unfortunately, in our implementation in the \emph{python}
programming language, the time consumption of even a tripple exchange
algorithm was found to be prohibitively large and we have not pursued
this any further.

The pairwise exchange method may reduce the number of students
assigned to their first preference. Although this is perfectly
allowed, we do build in a small bias towards rank one: Besides
exchanges that decrease the average rank, we also consider exchanges
that leave the average rank invariant, but for which the minimum of
the rank of the two students after the exchange is smaller than
before. That is, we prefer an assignment with ranks one and three to an
assignment with ranks two and two, etc.

Our pairwise exchange algorithm thus becomes:
\begin{enumerate}
\item order the pupils in decreasing rank according to the original
  solution;
\item starting from the first pupil, labeled $i$, consider an exchange
  with all other pupils, labeled $j$;
\item if the change in the average rank is smaller than zero, or if it
  is equal to zero but the minimum of the ranks of the two students
  becomes smaller, make the exchange;
\item continue until exchanges of all pairs of pupils have been
  considered.
\end{enumerate}
We have found that it does not improve the performance of the exchange
algorithm on our scenarios if the pupils are sorted again after every
exchange. However, we do 'restart' the loop on pupil $j$ after a
successful exchange. By running the algorithm more than once we have
verified that the algoritm effectively converges in one iteration.

\begin{figure}[htb]
  \centerline{\includegraphics[width=\textwidth]{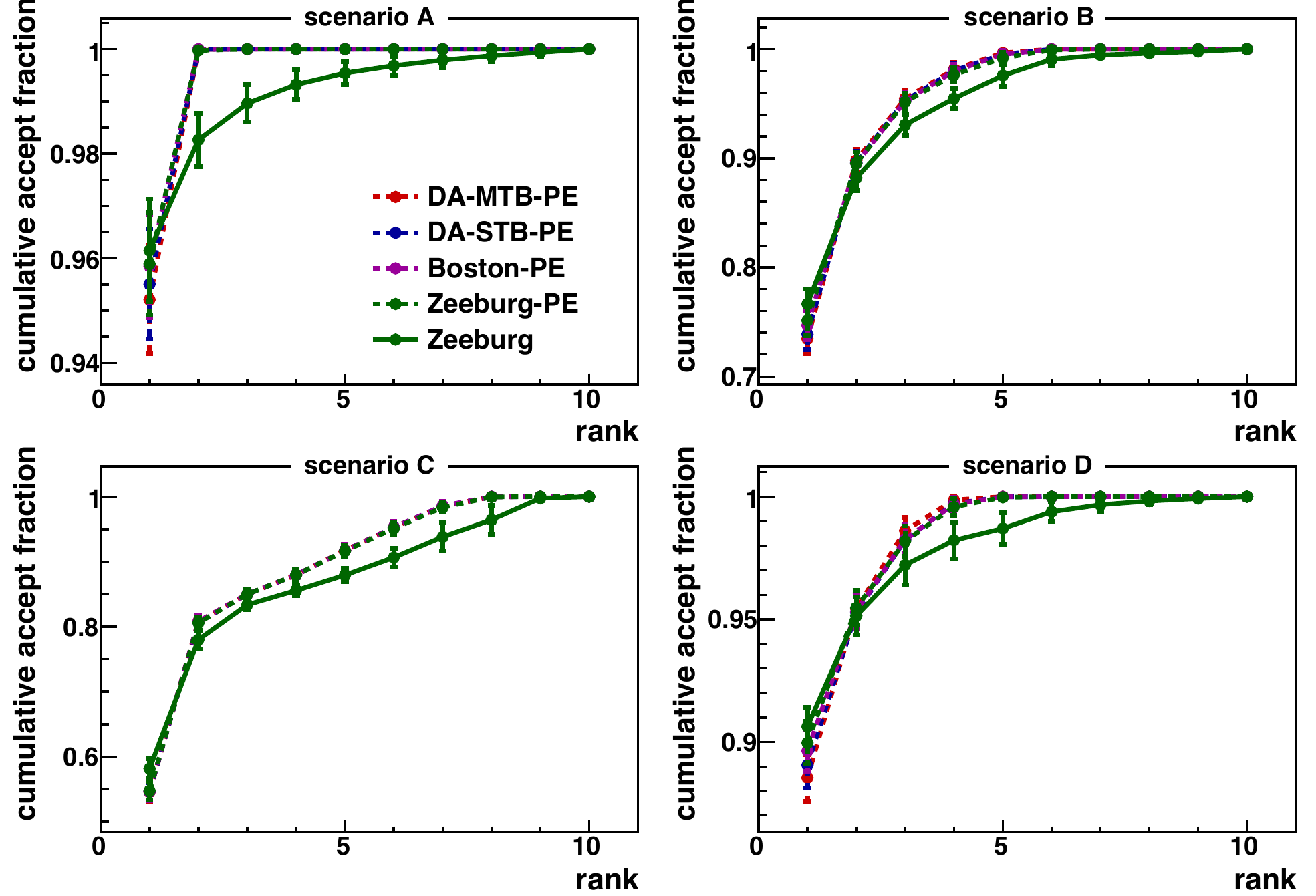}}
  \caption{Average cumulative acceptance functions for a 1000
    experiments in the four considered scenarions and for the DA-MTB,
    DA-STB, Boston and Zeeburg algorithm after the pairwise exchange
    (PE) algorithm. The Zeeburg algorithm without PE is included for
    the comparison. The vertical error bars correspond to the RMS of
    the variation between experiments.}
  \label{fig:acceptancecurvesPE}
\end{figure}

Figure~\ref{fig:acceptancecurvesPE} shows the cumulative acceptance
functions for all four algorithms after the PE algorithm is
applied. It is both interesting and reassuring that the curves depend
little on which algorithm was used to provide the solution that the PE
starts from. The PE algorithm can be successfully applied to improve
the inefficiency of any of the tested algorithms to about the same
level. This is also indicated by the average rank $Q$ shown in
Tab.~\ref{tab:averagerank} for all scenarios and for all algorithms
before and after the pairwise exchange.

\begin{figure}[htb]
  \centerline{\includegraphics[width=\textwidth]{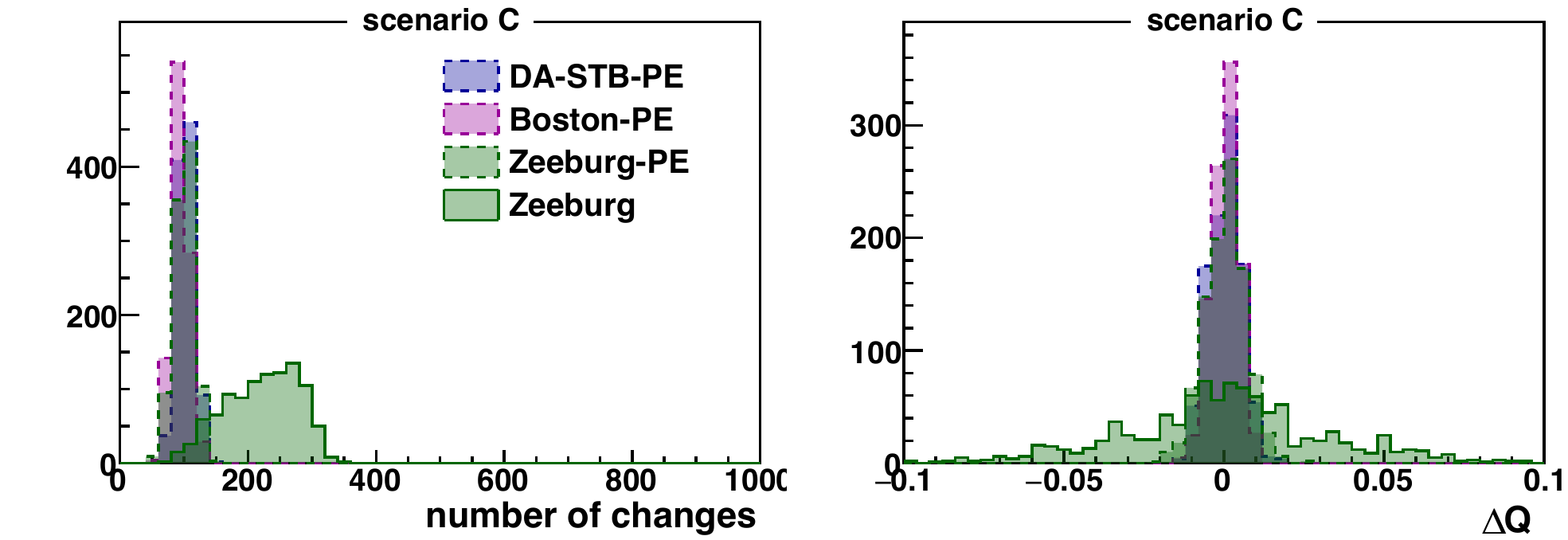}}
  \caption{Distributions for the number of students changing school
    (left) and the change in the average rank (right) after two consecutive
    calls to the algorithm in scenario C for 1000 experiments.}
  \label{fig:schoolswapsPE}
\end{figure}

\begin{table}[htb]
  \centerline{\small
    \begin{tabular}{|l|c|c|c|c|}
      \hline
      &  A  &  B  &  C  &  D \\
      \hline
      DA-MTB  & $ 1.14 \pm 0.05 $   & $ 3.03 \pm 0.13 $   & $ 3.96 \pm 0.09 $   & $ 1.79 \pm 0.06 $  \\
      DA-STB  & $ 1.11 \pm 0.03 $   & $ 2.17 \pm 0.06 $   & $ 2.76 \pm 0.05 $   & $ 1.45 \pm 0.04 $  \\
      Boston  & $ 1.10 \pm 0.03 $   & $ 1.95 \pm 0.05 $   & $ 2.53 \pm 0.04 $   & $ 1.41 \pm 0.03 $  \\
      Zeeburg  & $ 1.08 \pm 0.02 $   & $ 1.51 \pm 0.04 $   & $ 2.26 \pm 0.07 $   & $ 1.21 \pm 0.03 $  \\
      \hline
      DA-MTB-PE  & $ 1.05 \pm 0.01 $   & $ 1.44 \pm 0.03 $   & $ 2.06 \pm 0.04 $   & $ 1.18 \pm 0.02 $  \\
      DA-STB-PE  & $ 1.04 \pm 0.01 $   & $ 1.44 \pm 0.03 $   & $ 2.06 \pm 0.04 $   & $ 1.18 \pm 0.02 $  \\
      Boston-PE  & $ 1.04 \pm 0.01 $   & $ 1.43 \pm 0.03 $   & $ 2.06 \pm 0.04 $   & $ 1.17 \pm 0.02 $  \\
      Zeeburg-PE  & $ 1.04 \pm 0.01 $   & $ 1.43 \pm 0.03 $   & $ 2.07 \pm 0.04 $   & $ 1.17 \pm 0.02 $  \\
      \hline
    \end{tabular}}
  \caption{Average rank $Q$ for the scenarios and algorithms
    discussed in the text for 1000 experiments. The quoted
    error is the standard deviation of the variation
    between experiments.}
  \label{tab:averagerank}
\end{table}

To illustrate the stability of the result
figure~\ref{fig:schoolswapsPE} (left) shows the fraction of students
changing schools for two independent sets of tie-breakers for a subset
of the DE improved algorithms in scenario C. Note that there is still
a large variation in the assignment. However, as illustrated in the
right figure, the solutions are actually very close in rank. We found
that most of the difference between the solution can be attributed to
pairs of students that have exchanged places such that the final
change is rank neutral, simply because the students have ranked the
two schools in the same way.

One may wonder how close to the optimal solution the result of the
pairwise exchange is. As the starting point is determined by the
random tie-breaker and only a finite set of pairwise exchanges is
tried, the result may correspond to a `local minimum' of the average
rank. As a different local minimum is obtained with a different
tie-breaker, one can try to assess the distance to the true minimum by
trying different random tie-breakers. We have compared the average rank
obtained with a single call to Boston plus PE to that obtained with a
pick of `the best of 10'. The difference was found to be small, of the
order of the variations seen on the right in
Fig.~\ref{fig:schoolswapsPE}. We did not study the asymptotic behaviour
in more detail but it seems that in practice the solution is close to
optimal.

\begin{figure}[htb]
  \centerline{\includegraphics[width=\textwidth]{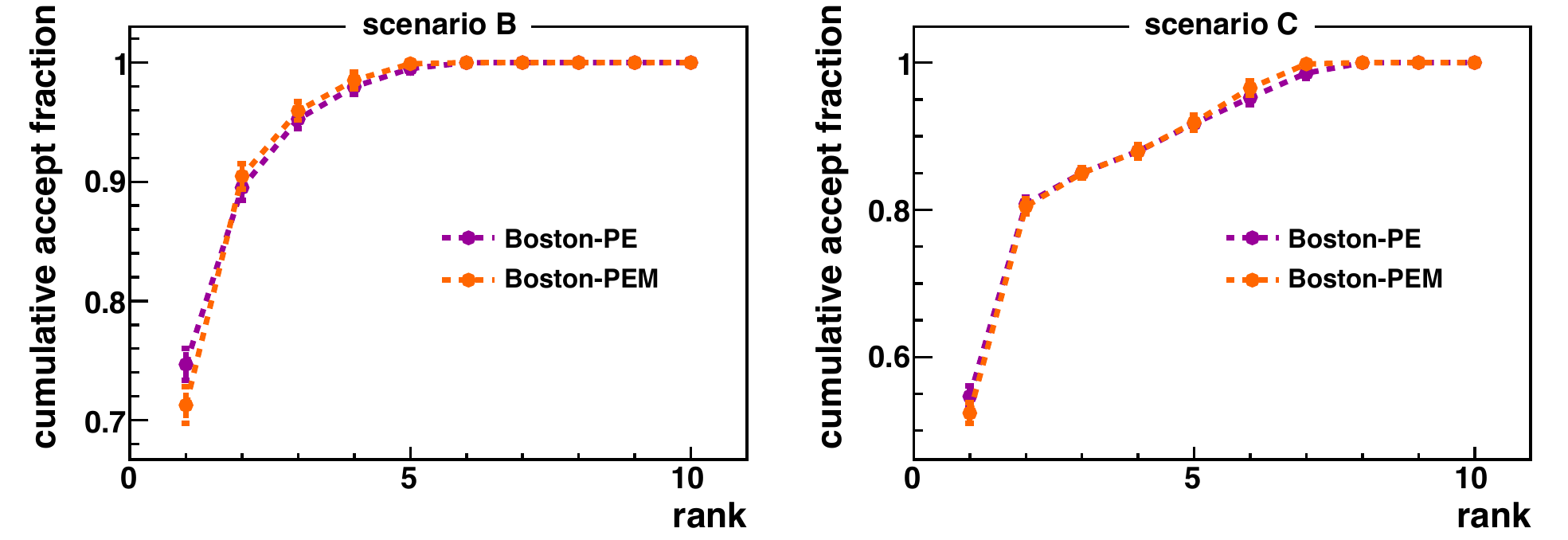}}
  \caption{Average cumulative acceptance functions for a 1000
    experiments in scenario B and C for Boston with the default
    pairwise exchange algorithm (PE) and for pairwise exchange with
    minimal variance (PEM).}
  \label{fig:PEalternatives}
\end{figure}

Finally, we have also compared the alternative option for dealing with
'neutral' exchanges, namely to choose the one with the smallest
maximum rank, rather than the smallest minimum rank. The former will
lead to a smaller variance of the rank distribution. The comparison
for Boston with PE is shown in Fig.~\ref{fig:PEalternatives}, where
the alternative is labeled with the abbreviation PEM. As expected, the
acceptance functions cross: the PE method gives slightly more results
with rank 1, but has slightly more tail. The average rank is
practically the same for both methods.

\subsection{Tests of strategy-proofness}

A matching method is strategy-proof if pupils do not benefit from
specifying a preference list different from their true ordinal
preference. It is not apriori clear what `benefit' means in this
context, since there is always a price to pay. As we shall see below,
students could apply a strategy that gives them a higher chance to get
their first preference, at the expense of having a higher change to
end up with a school that ranks low on their list; or they could aim
to increase the chance to get within their top three, by ranking their
actual first choice lower. Therefore, one may argue that determining a
strategy is just a cost-benefit analysis that individual pupils should
be allowed to make.

The main reason that we should worry about strategy-proofness anyway
is because pupils that do not apply a strategy may be harmed by the
behaviour of the strategists. This leads to a form of inequality as
the background of students and parents influences their ability to
understand the consequences of different strategies. In the following
we test the effect of two simple selection strategies in our
simulations.

It should be emphasised that for a subset of students the current
system in Amsterdam is already not strategy-proof for any matching
algorithm. The reason is that some schools give preference to students
that either have brothers or sisters at the same school, or that
attended a certain type of primary school. As this preference is only
given if students rank the school first, it is an incentive to put the
school at the first place, even if it is not actually the first
preference.

To investigate strategy-proofness we can compare the efficiency of the
matching algorithms for students that apply different kind of
strategies. We have found that, in practice, it is not that simple to
define a popularity scenario and a ranking strategy that actually lead
to a benefit for strategic students. After some trial and error, we
have come up with scenario C: two schools that are so popular that
mosts student will rank them as one and two, and two schools that are
so unpopular that they are almost always at the bottom of the list.

\begin{figure}[htb]
  \centerline{\includegraphics[width=\textwidth]{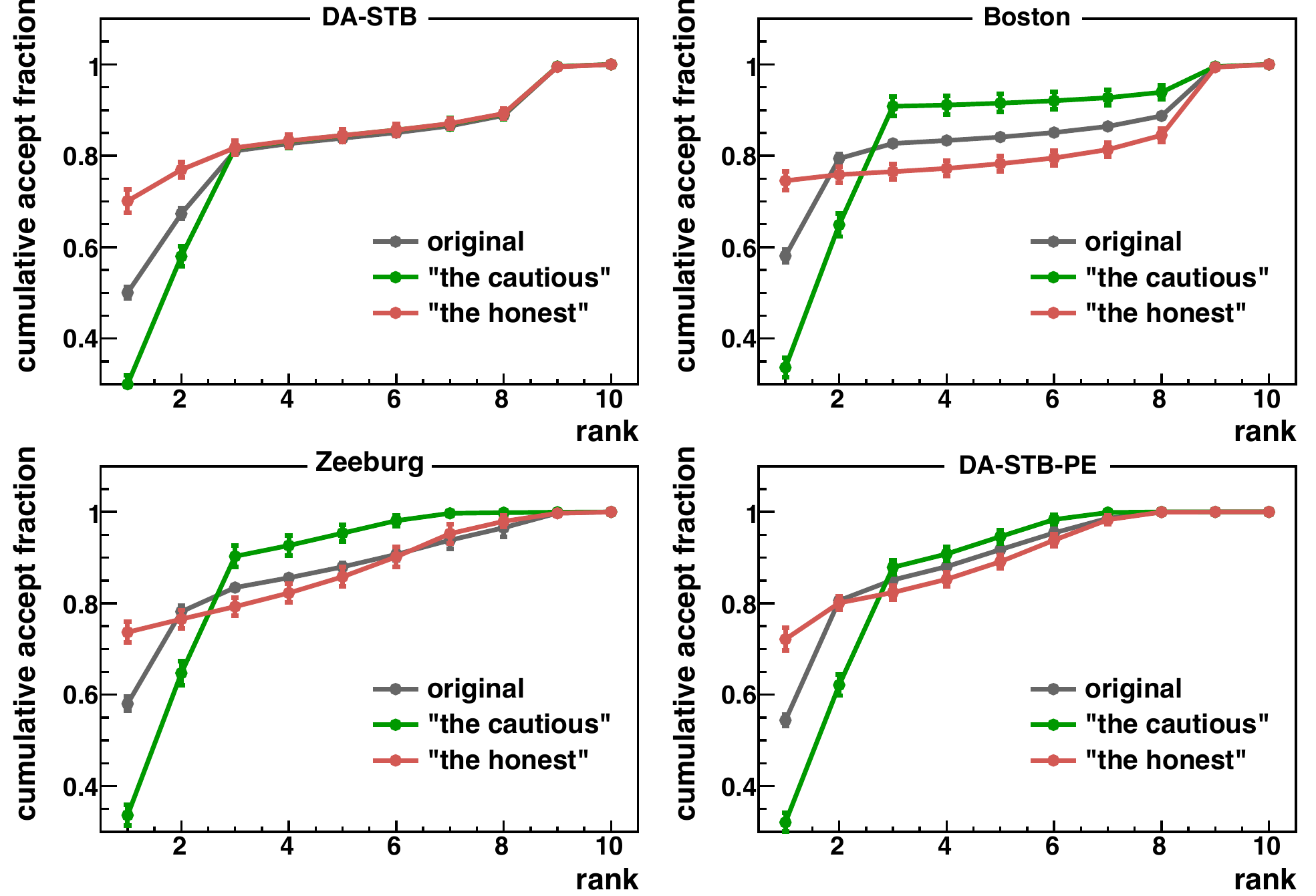}}
  \caption{Average cumulative acceptance as function of the
    \emph{true} rank with and without 50\% of the students following
    the `cautious' strategy described in the text for 100 experiments
    according to popularity scenario C.}
  \label{fig:strategy}
\end{figure}

In this scenario strategic students can try to evade the unpopular
school by putting one of the less unfavourable schools in their top
two. To keep the implementation generic the actual applied strategy is
that students re-order their true top three according to the known
average popularity
(table~\ref{tab:scenarios}). Figure~\ref{fig:strategy} shows the
effect on the acceptance curves in a simulation of scenario C with
50\% of the students applying this strategy. Note that the cumulative
acceptance is given as function of the \emph{true} rank, not the rank
that the strategic student provided. The students applying a strategy
are called `cautious', while the remaining students in the sample are
`honest'. For reference also the original curves with only honest
students are shown.

As expected, the DA algorithm with a random tie-breaker (be it STB or
MTB) is indeed strategy-proof: students applying the cautious strategy
are worse of than honest students, so applying a strategy makes no
sense. The Boston algorithm is not strategy-proof in this scenario:
although the cautious loose on their top one and two ranking, they
beat their victims in the top three and beyond. The Zeeburg algorithm
and any algorithm combined with pairwise-exchange optimisation are not
strategy-proof either. Interesting enough, in this scenario, they seem
to be more strategy-proof than Boston, even though they are more
efficient. This shows that efficiency is not directly coupled to
strategy-proofness.

In any case, it is important to note that in this scenario the victims
are not worse off with any of the improved algorithms than they are
with DA-STB. This can be seen by comparing the `original' curve for
DA-STB with the `honest' curve in DA-STB-PE.  The costs of DA's
strategy-proofness is simply too high to compensate for the
inefficiency caused by the lack of strategy-proofness in the other
algorithms.

\begin{figure}[htb]
  \centerline{\includegraphics[width=\textwidth]{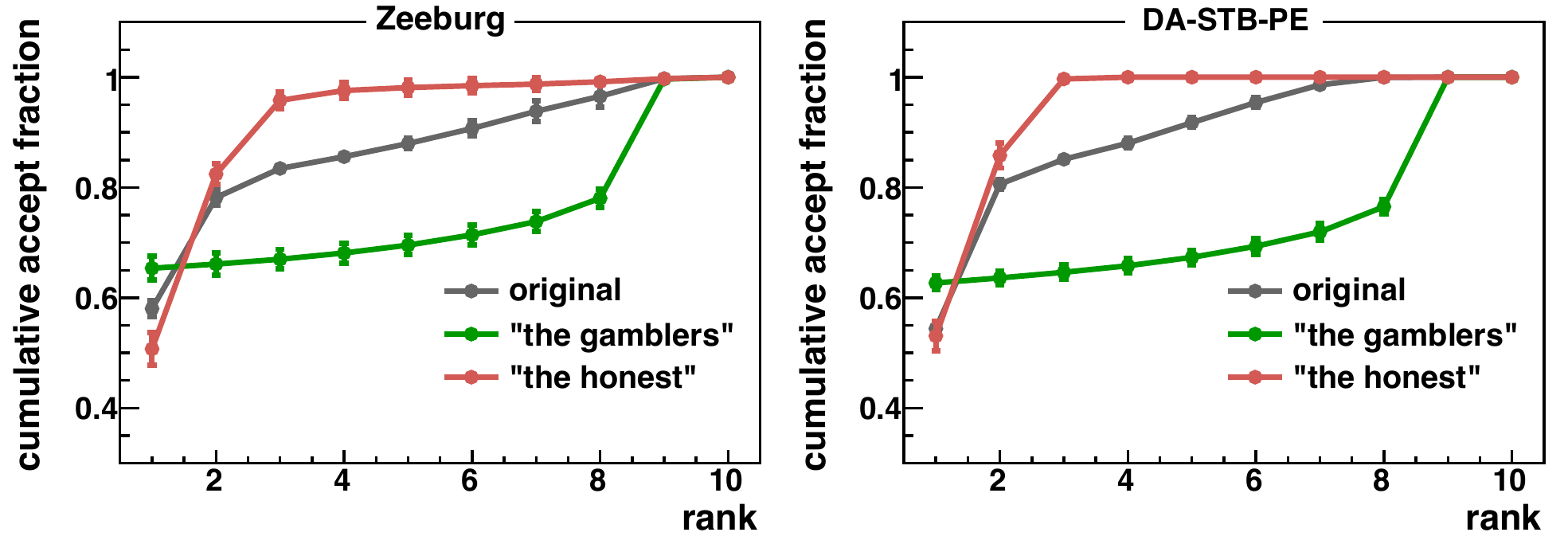}}
  \caption{Average cumulative acceptance as function of the
    \emph{true} rank with and without 50\% of the students following
    the `gambling' strategy described in the text for 100 experiments
    according to popularity scenario C.}
  \label{fig:strategyG}
\end{figure}

Students could also apply a strategy that increases their chances to
get assigned to their first choice by exploiting that some of the
algorithms effectively give higher preferences to students that are
more difficult to place at another school. This holds in particular
for the Zeeburg and PE algorithm. We implement this strategy by
keeping the first choice as is as, but rank the remaining schools in
order of decreasing popularity. We call this the `gambling' strategy
as these students give up on anything but their first choice. The
result is shown for Zeeburg and DA-STB-PE in
figure~\ref{fig:strategyG} for a scenario with 50\% gamblers. Clearly,
the gamblers manage to profit from their strategy as the fraction of
them that gets a rank one assignment is larger than for the 'honest
only' scenario. However, it also shows that the effect on the
remaining students is small. Those students are still better of with
the improved algorithm than with the original strategy-proof
algorithm.

\begin{table}[htb]
  \centerline{\small
    \begin{tabular}{|l||c|c||c|c|}
      \hline
      {} & \multicolumn{2}{c||}{cautious} & \multicolumn{2}{c|}{gambling}\\
      \hline
      {} & strategists & honest & strategists & honest \\
      \hline
      Boston & $ 2.50 $ & $ 2.73 $ & $ 3.68 $ & $ 1.66 $ \\
      Zeeburg & $ 2.26 $ & $ 2.20 $ & $ 3.41 $ & $ 1.79 $ \\
      Boston-PE & $ 2.33 $ & $ 2.00 $ & $ 3.58 $ & $ 1.61 $ \\
      Zeeburg-PE & $ 2.31 $ & $ 2.03 $ & $ 3.55 $ & $ 1.63 $ \\
      DA-STB-PE & $ 2.34 $ & $ 1.99 $ & $ 3.58 $ & $ 1.61 $ \\
      \hline
    \end{tabular}
  }
  \caption{Average (true) rank $Q$ for strategic and honest pupils measured over 100 experiments in scenario
    C with either 50\% cautious strategists (left) or 50\% gambling
    strategists (right).}
  \label{tab:averagerankstrategising}
\end{table}

Table~\ref{tab:averagerankstrategising} shows the average rank
obtained for strategic and honest pupils in the scenarios above. One
clearly observes the price the strategists pay: as they do not provide
their true preferences, their average rank is usually higher than for
the honest students. Comparing to Tab.~\ref{tab:averagerank} we find
that in terms of the average rank the honest do not score worse in
Zeeburg and PE algorithms than they did without the strategists.

In the other scenarios (A, B and D) the negative effect on honest
students as a result of cautious and gambling strategies was found to
be insignificant. That does not mean that there do not exist scenarios
in which honest students are better off with a strategy-proof
inefficient algorithm as DA-STB. However, it illustrates that in
practice such scenarios may be rare.

\section{Practical considerations and other discussion points}

\subsection{School preferences}

Some of the schools in Amsterdam may give a subset of pupils a
preference over others, for example because elder siblings attend the
school. In order to respect these constraints, they need to be built
into the tie-breaker at the school. It is not easy to use such
constraint in the Zeeburg algorithm. The easiest solution is to deal
with this subset of students first, and use the matching algorithms
only for the students that remain.

\subsection{Incomplete preference lists}

Above we have simulated a situation in which all pupils submit an
ordered list that contains all schools. In Amsterdam, pupils do not
need to hand in a full list: they may hand in a list with just one
school. If a pupil cannot be assigned to that school in Boston or
DA, the consequence is that the pupil needs to participate in a second
round, in which only schools participate that still have places left.

Clearly, this has consequences for the implementation of the
algorithms. For instance, the Zeeburg and pairwise exchange cannot be
applied in a fair way unless all preference lists are complete, as
students may on purpose hand in lists that do not contain less popular
schools.

One practical solution is to complete the preference lists. They could
be completed deterministically as follows: once the preference lists
are available, schools are ranked by popularity. Every pupil
preference lists is completed with the missing schools in order of
increasing popularity. This is a clear motivation for students to hand
in a long preference list.

\subsection{School types}

In contrast to many other countries in the world, the school system in
the Netherlands differentiates the level of education directly at the
start of secondary education. The level appropriate for a pupil's
secondary education is determined by the teachers at the primary
school based on scores to standard tests performed during the pupil's
primary school career. The proposed level is called the \emph{advice}.

There are roughly four `levels' of education. In theory, this just
splits the matching in four independent parts. In practice, it is not
that simple. First, students may be given a mixed advice. Second, many
schools offer transition classes for the first or second year that
combine more than one level.

This complication is not a show-stopper, however. If the matching can
be applied with Boston or DA, then the pairwise exchange algorithm can
be applied a such, as long as it only exchanges students that have the
same school advice.

\subsection{Simplicity}

An important property of a suitable matching algorithm is that it is
sufficiently simple that it can be both easily be explained and
unambiguously described and implemented. In this respect the Zeeburg
algorithm is perhaps a bordercase. However, the pairwise exchange
algorithm certainly qualifies as simple.

\subsection{Alternative optimisation criterion}

The pairwise exchange method optimised the average rank $Q$. In the
optimisation the difference between rank 8 and 9 is the same as
between rank 1 and 2. However, pupils probably care less about the
order in the tail, than the order of their top ranked schools. This
was the main reason that we preferred the PE method with larger
variance over the one with minimal variance (PEM).

Still, one may wonder if alternative definitions of $Q$, for example
as a power-law $Q \propto \sum_i r_i^\alpha$ with $\alpha<1$, would not
lead to a solution that better reflects the cardinal preferences of
the pupils. Inevitably, this will lead to a larger tail in the rank
distribution. We have not further investigated this.

\section{Conclusions}

As was known long before it was introduced in Amsterdam, the DA
algorithm is not a particularly efficient solution to the college
admission problem with
indifference~\cite{Abdulkadiroglu:2008,Abdulkadiroglu:2009}, as it was
developed for a two-sided market problem with preferences on both
sides. The Boston algorithm better respects the students
preferences. Other algorithms, such as the Zeeburg algorithm and the
pairwise exchange optimisation introduced here, perform even better,
in a variety of simple scenarios. The reason is that the sensitivity
to the tie-breaker, the lottery tickets of the pupils, is
significantly smaller in these alternatives.

The inefficiency of the DA algorithms with random tie-breakers is the
cost of strategy-proofness~\cite{Abdulkadiroglu:2008}. The more efficient
algorithms are not strategy-proof. However, in the considered
scenarios the costs of the lack of strategy-proofness is smaller than
the costs of the inefficiency of DA: Even students that do not apply a
strategy are better of with the non-strategy-proof
algorithms. Therefore, it seems hard to maintain strategy-proofness as
a requirement of the matching algorithm.

To understand whether or not these conclusions hold in more realistic
scenarios, an analysis like the one in~\cite{Oosterbeek:2015} will
need to be performed.  However, based on the current results we
strongly advise local authorities to reconsider their choice for DA in
school matching. The most simple way to `fix' the algorithm is to
augment it with the pairwise exchange algorithm that we
described. This algorithm is simple and suffers little from the
practical limitations discussed above. We believe that by applying
this method, the results of the matching will be significantly more in
line with the students preferences.

From personel experience we know that pupils and their parents spend a
lot of effort to prioritise the schools in Amsterdam. If these efforts
are taken seriously, random number should play a minimal role in the
matching.

\section{Acknowledgements}

The author would like to thank Dr.~Paola Grosso, Dr.~Edward
Hulsbergen, Rola Hulsbergen-Paanakker, Prof.~Dr.~Olga Igonkina,
Prof.~Dr.~Gerhard Raven, Hartmut Samtleben and Dr.~Wouter Verkerke,
for stimulating discussions and corrections to this write-up.

\appendix

\section{The Zeeburg algorithm}
\label{app:zeeburg}

The DA and Boston algorithms rely on a random tie-breaker that
effectively describes the school preferences. We have implemented
another algorithm in which the number of ties broken by the random
tie-breaker is minimised. The algorithm works as follows:

\begin{enumerate}
\item For every school keep track of 
  \begin{enumerate}[(i.)]
  \item the number of vacant places at the school;
  \item the rank of the current queue;
  \item the pupils in the queue.
  \end{enumerate}
  In addition keep track of the list of completed student-school
  matches.  This defines the \emph{state} of the algorithm.
\item Sort pupils according to a single random tie-breaker. Set the
  rank of the queue of every school to one and assign the number of
  vacant places. Line up pupils in the queue of their favourite
  school. This populates the queue in every school and completes the
  \emph{initial} state of the algorithm;
\item Now run the following loop:
  \begin{enumerate}[(a.)] 
  \item select a school that can entirely admit the queue of its
    current rank. If there is more than one such school, select the
    school with the queue with the smallest rank. If there is more
    than one such queue, select the school for which the number of
    places remaining after accepting all pupils in the queue is the
    smallest;
  \item for this queue, accept all pupils. Remove these pupils in
    every other queue that they appear. (Initially, pupils appear in
    only one queue, but this changes while the algorithm is running.)
    Reduce the number of vacant places according to the number of
    newly accepted students. If there are places left at the school,
    increase the rank of the queue, and line up all pupils that have
    not yet found a place and that rank the school according to the
    rank of the queue. (These students will now be in more than one
    queue);
  \item repeat until the condition under (3a.) can no longer be
    satisfied for any school.
  \end{enumerate}
\item Apply the tie-breaker to force a decision on one of the queues:
  \begin{enumerate}[(a.)]
  \item select the queue with the smallest rank in a school that is
    not yet full. If there is more than one such queue, select the
    queue that has the smallest overflow (that is, for which the
    length of the queue minus the number of available places is
    minimal);
  \item accept pupils from the start of this queue until the school is
    full. Remove accepted pupils from other queues;
  \end{enumerate}
\item Repeat steps 3 and 4 until all pupils have been accepted.
\end{enumerate}


\end{document}